\documentstyle[prb,aps,epsf]{revtex}

\begin{document}
\draft
\tightenlines

\title{Dynamical Charge Susceptibility of the Spinless Falicov-Kimball Model.}

\author{ J. K. Freericks and P. Miller }

\address{
Department of Physics, Georgetown University, \\
 Washington, D.C. 20057-0995, U.S.A. }
\maketitle
\begin{abstract}
An exact solution is presented for the frequency-dependent
charge susceptibility of the spinless Falicov-Kimball model by
using dynamical mean-field theory.  We develop a nontrivial
application of the Baym-Kadanoff ``conserving approximation'' formalism
to exactly determine the frequency-dependent vertex function (which turns
out to assume a particularly simple form). We show
how the static and dynamic susceptibilities are decoupled in this model
and how the dynamic susceptibility generically does not show any signal
of the low-temperature charge-density-wave phase transition.  We also
examine the temperature evolution of the dynamic charge susceptibility
for the special case of half-filling.
\end{abstract}

\pacs{}

\section{Introduction}

The Falicov-Kimball model\cite{falicov_kimball} was introduced in 1969 as a
thermodynamic model for metal-insulator transitions in systems that have two
different types of electrons---itinerant conduction electrons and localized
$f$-electrons.  Twenty years later, Brandt and Mielsch\cite{brandt_mielsch}
solved the Falicov-Kimball model exactly in the limit of infinite dimensions
by using dynamical mean-field theory and the Baym-Kadanoff formalism.
They also showed how to calculate static susceptibilities and found an 
Ising-like phase transition to a chessboard charge-density-wave phase at
half-filling.  Freericks\cite{freericks} then showed that the system also 
displayed incommensurate order and phase separation at other fillings.

The Hamiltonian of the spinless Falicov-Kimball model\cite{falicov_kimball} is
\begin{equation}
\hat 
H =-\frac{t^*}{2\sqrt{d}}\sum_{\langle i,j\rangle}d_i^{\dagger}d_j +\epsilon_f
\sum_i w_i-\mu\sum_i(d_i^{\dagger}d_i+w_i)+U\sum_id_i^{\dagger}d_iw_i,
\label{eq: ham}
\end{equation}
where $d_i^{\dagger}$ $(d_i)$ is the spinless conduction electron creation
(annihilation) operator at lattice site $i$ and $w_i=0$ or 1 is a classical
variable corresponding to the localized $f$-electron number at site $i$.  The
hopping matrix between nearest neighbors $\langle i,j\rangle$ (on a hypercubic
lattice in $d$-dimensions, with $d\rightarrow\infty$)
is $-t^*/(2\sqrt{d})$ with $t^*$ chosen as our energy unit,
$\epsilon_f$ is the localized electron level, $\mu$ is the chemical potential
and $U$ is the mutual electron repulsion when a conduction electron and a 
localized $f$-electron both occupy the same lattice site.

In this contribution we examine the $d$-electron dynamical charge susceptibility
[where $n_i(\tau)=\exp(\tau \hat H )n_i(0)\exp(-\tau \hat H )$ and $n_i=
d_i^{\dagger}d_i$]
\begin{equation}
\chi({\bf q},i\nu_l)={\rm Tr}T_{\tau}\sum_{j,k}\int_0^{\beta}{\mathrm d}\tau 
e^{i\nu_l\tau}e^{i{\bf q}\cdot({\bf R_j}-{\bf R_k})}\left [
\frac{\langle e^{-\beta \hat H}n_j(\tau)n_k(0)\rangle}{Z}-
\frac{\langle e^{-\beta \hat H}n_j\rangle}{Z}
\frac{\langle e^{-\beta \hat H}n_k\rangle}{Z}\right ],
\label{eq: chidef}
\end{equation}
and its analytic continuation to the real-frequency axis.  Here 
$i\nu_l=2i\pi Tl$
is the Bosonic Matsubara frequency, $Z$ is the partition function, and 
${\bf R_j}$ is the position operator for lattice site $j$.  The Falicov-Kimball
model is the simplest many-body problem that has nontrivial dynamics.  The
model is simple enough that an exact solution can be found for the charge
susceptibility, but is complicated enough to show many-body effects.
Shvaika\cite{shvaika} has shown how to determine the lattice charge 
susceptibility from a diagrammatic analysis of the atomic problem.  Here we
provide an alternate derivation, stressing the Baym-Kadanoff approach (which
produces a much simpler form for the vertex), and we
provide numerical results to complement the formalism.

We begin with a discussion of some special symmetries of the Falicov-Kimball
model.  Since the total conduction electron number is a conserved quantity,
we find that $\sum_jn_j(\tau)=\sum_jn_j(0)$ has no $\tau$-dependence.  Hence,
the uniform charge susceptibility ({\bf q}=0) vanishes for all nonzero
frequencies.  Similarly, the local $f$-occupation is also conserved [$w_i(\tau)
=w_i(0)$], since $w_i$ commutes with the lattice Hamiltonian.  This implies that
both the $ff$-charge susceptibility and the mixed $df$-charge susceptibility
have no $\tau$-dependence and therefore 
are static, with no frequency dependence.  Furthermore, a simple
application of the chain rule and the definition of each of the
susceptibilities as a derivative of the corresponding average electron
filling with respect to an external field, shows that there is a discontinuity
in the $dd$-charge susceptibility at zero frequency\cite{shvaika} due to the
coupling of the $d$ and $f$-electrons.  This implies that there is a decoupling
between the static and dynamic $dd$-charge susceptibility!

The solution of the Falicov-Kimball model has been outlined in detail
elsewhere\cite{brandt_mielsch,freericks,freericks_zlatic}.  Here we summarize
the main points to establish our notation.  The local Green's function at
the Fermionic Matsubara frequency $i\omega_n=i\pi T(2n+1)$ is defined by
\begin{equation}
G_n=G(i\omega_n)=-{\rm Tr_d Tr_f}T_{\tau}\int_{0}^{\beta}{\mathrm d}\tau 
e^{i\omega_n \tau}\frac{\langle 
e^{-\beta \hat H_{at}} d(\tau)d^{\dagger}(0)S(\lambda) \rangle}{Z},
\label{eq: gdef}
\end{equation}
with 
\begin{equation}
Z=Z_0(\mu)+e^{-\beta(\epsilon_f-\mu)}Z_0(\mu-U),
\label{eq: zdef}
\end{equation}
the atomic partition function expressed in terms of
\begin{equation}
Z_0(\mu)={\rm Tr_d}\langle e^{-\beta \hat H_0}S(\lambda)\rangle,\quad 
\hat H_0=-\mu d^{\dagger}d,
\label{eq: z0def}
\end{equation}
In the above equations, the atomic Hamiltonian
$\hat H_{at}$ is the Hamiltonian of Eq.~(\ref{eq: ham})
restricted to one site, with $t^*=0$, and all time dependence is with respect
to this atomic Hamiltonian.  The evolution operator $S(\lambda)$ 
satisfies
\begin{equation}
S(\lambda)=\exp[-\int_0^{\beta}{\mathrm d}\tau\int_0^{\beta}{\mathrm
d}\tau^{\prime} d^{\dagger}(\tau)\lambda(\tau-\tau^{\prime})d(\tau^{\prime})],
\label{eq: sdef}
\end{equation}
with $\lambda(\tau-\tau^{\prime})$
a time-dependent atomic field adjusted to make
the atomic Green's function equal to the local lattice Green's function.  We
define an effective medium by
\begin{equation}
G_0^{-1}(i\omega_n)=G_n^{-1}+\Sigma_n=i\omega_n+\mu-\lambda_n,
\label{eq: g0def}
\end{equation}
with $\Sigma_n$ the local self-energy and $\lambda_n$ the Fourier transform of
$\lambda(\tau)$.  The trace in Eq.~(\ref{eq: gdef}) can be evaluated 
directly to yield
\begin{equation}
G_n=w_0G_0(i\omega_n)+w_1[G_0^{-1}(i\omega_n)-U]^{-1},
\label{eq: gatomic}
\end{equation}
with $w_0=1-w_1$ and
\begin{equation}
w_1=\exp[-\beta(\epsilon_f-\mu)]Z_0(\mu-U)/Z.
\label{eq: w1def}
\end{equation}
The self-consistency relation needed to determine $\lambda_n$ and hence $G_n$
is to equate the local lattice Green's function to the atomic Green's function
via
\begin{equation}
G_n=\int_{-\infty}^{\infty}d\epsilon\frac{\rho(\epsilon)}
{i\omega_n+\mu-\Sigma_n-\epsilon},
\label{eq: glat}
\end{equation}
with $\rho(\epsilon)=\exp(-\epsilon^2)/\sqrt{\pi}$ the noninteracting density
of states for the infinite-dimensional hypercubic lattice.

The iterative algorithm to solve for $G_n$ starts with $\Sigma_n=0$.  Then 
Eq.~(\ref{eq: glat}) is used to find $G_n$, Eq.~(\ref{eq: g0def}) is employed
to extract the effective medium, Eq.~(\ref{eq: gatomic}) is used to find a
new local Green's function, and then Eq.~(\ref{eq: g0def}) is used to find
the new self-energy.  The algorithm is then repeated until it converges,
which usually requires only about a dozen or so iterations.  In this 
contribution, we examine the half-filled case $\rho_d=\sum_i\langle n_i\rangle
/N =0.5$ and $\rho_f=\sum_i\langle w_i\rangle /N=0.5$, which corresponds to
$\mu=U/2$ and $\epsilon_f=-U/2$.

In the following section we illustrate how the Baym-Kadanoff formalism can be
used to determine the dynamical charge susceptibility.  Numerical results 
at half filling are presented in Section III and conclusions in Section IV.

\section{Baym-Kadanoff formalism}

The momentum-dependent susceptibility satisfies the following Dyson's equation:
\begin{eqnarray}
\chi^{dd}({\bf q},i\omega_m,i\omega_n;i\nu_l)&=&\chi_0^{dd}({\bf q},i\omega_m;
i\nu_l)\delta_{mn}\cr
&-&T\sum_{n^{\prime}}\chi_0^{dd}({\bf q},i\omega_m;i\nu_l)\Gamma(i\omega_m,
i\omega_{n^{\prime}};i\nu_l)
\chi^{dd}({\bf q},i\omega_{n^{\prime}},i\omega_n;i\nu_l),
\label{eq: chidyson}
\end{eqnarray}
and the full susceptibility is found by summing over the Fermionic
Matsubara frequencies $\chi^{dd}({\bf q},i\nu_l)=T\sum_{mn}\chi^{dd}
({\bf q},i\omega_m,i\omega_n;i\nu_l)$.  In Eq.~(\ref{eq: chidyson}) the
bare susceptibility satisfies
\begin{eqnarray}
\chi^{dd}_0(X,i\omega_m;i\nu_l)&=&-\sum_n\sum_{\bf k}G_m({\bf k})G_{m+l}({\bf k}
+{\bf q}),\cr
&=&-\frac{1}{\sqrt{1-X^2}}\int_{-\infty}^{\infty}d\epsilon\frac{\rho(\epsilon)}
{i\omega_m+\mu-\Sigma_m-\epsilon}F_{\infty}\left ( \frac{i\omega_{m+l}+\mu
-\Sigma_{m+l}-X\epsilon}{\sqrt{1-X^2}}\right ),
\label{eq: chi0def}
\end{eqnarray}
with $X({\bf q})=\lim_{d\rightarrow\infty}\sum_{i=1}^d\cos{\bf q_i}/d$ 
describing all of the
momentum dependence of the susceptibility\cite{mueller-hartmann,brandt_mielsch}
and
\begin{equation}	
F_{\infty}(z)=\int_{-\infty}^{\infty}d\epsilon\frac{\rho(\epsilon)}{z-\epsilon},
\label{eq: fdef}
\end{equation}
the Hilbert transform of the noninteracting density of states.

The dynamical susceptibility simplifies in three cases: the chessboard case
[${\bf q}=(\pi,\pi,...,\pi),$ $X=-1$] where 
\begin{equation}
\chi_0^{dd}(-1,i\omega_m;i\nu_l)=-\frac
{G_m+G_{m+l}}{i\omega_m+i\omega_{m+l}+2\mu-\Sigma_m-\Sigma_{m+l}};
\label{eq: chi0_-1}
\end{equation}
the local case ($X=0$) where
\begin{equation}
\chi_0^{dd}(0,i\omega_m;i\nu_l)=-G_mG_{m+l};
\label{eq: chi0_0}
\end{equation}
and the uniform case [${\bf q}=(0,0,...,0),$ $X=1$] where
\begin{equation}
\chi_0^{dd}(1,i\omega_m;i\nu_l)=-\frac{G_m-G_{m+l}}
{i\nu_l+\Sigma_m-\Sigma_{m+l}}.
\label{eq: chi0_1}
\end{equation}
We will only be interested in these three simpler cases here.  Note that in
instances where the denominators of Eqs.~(\ref{eq: chi0_-1}) and 
(\ref{eq: chi0_1}) vanish, the susceptibility is evaluated by l'H\^opital's
rule.

The calculation of the full susceptibility requires the local irreducible
vertex function.
The Baym-Kadanoff approach~\cite{Baym1,Baym2} 
solves for the irreducible vertex function in a 
manner which guarantees that an approximation scheme maintains the 
conservation laws of the Hamiltonian. 
The procedure requires that an approximate self-energy, 
$\Sigma({\bf r_{1}},\tau_{1}, {\bf r_{2}},\tau_{2})$ be given by the 
functional derivative of a free-energy 
functional, $\Phi$, with respect to the full Green's function, 
$G({\bf r_{1}},\tau_{1}, {\bf r_{2}},\tau_{2})$. That is:
\begin{equation}
\Sigma({\bf r_{1}},\tau_{1}, {\bf r_{2}},\tau_{2}) = 
\frac{ \delta \Phi }{\delta G({\bf r_{1}},\tau_{1}, {\bf r_{2}},\tau_{2}) },
\label{eq: sigma_func}
\end{equation}
and then 
differentiation of the self-energy with respect to the Green's function  
produces the irreducible vertex function
\begin{equation} 
\Gamma({\bf r_{1}},\tau_{1}, {\bf r_{2}},\tau_{2};
{\bf r_{3}},\tau_{3}, {\bf r_{3}},\tau_{3})
=\frac{1}{T} \frac{ \delta \Sigma({\bf r_{1}},\tau_{1}, {\bf r_{2}},\tau_{2}) }
{\delta G({\bf r_{3}},\tau_{3}, {\bf r_{3}},\tau_{3})}.
\label{eq: gamma_func}
\end{equation}
Of course, this technique can also be used for an exact solution as we do
here.

Using the dynamical-mean-field approximation for the Falicov-Kimball model in 
a spatially invariant system yields an exact expression for the self-energy
$\Sigma=G_0^{-1}-G^{-1}$ [as seen in Eq.~(\ref{eq: g0def})]
and we do not need to find the appropriate free-energy functional $\Phi$. 
Our strategy is to calculate the Green's function of the effective medium, 
$G_{0}(\tau_{1},\tau_{2})$, when there is an additional time-dependent field, 
$\chi(\tau)$, which couples to the charge density, 
${d}(\tau){d}^{\dagger}(\tau)$. 
[{\it i.e.} we add 
$\int^{\beta}_{0} {\mathrm d} \tau \chi(\tau) 
{d}^{\dagger}(\tau){d}(\tau)$ 
to the exponent of the evolution operator in Eq.~(\ref{eq: sdef})]. 
This time-dependent field removes time-translational 
invariance from the system, so the Green's functions now depend on $\tau_{1}$ 
and $\tau_{2}$ separately. 
Using $G_{0}(\tau_{1},\tau_{2})$, we evaluate the 
self-energy as an explicit function of the full Green's function, 
$G(\tau_{1},\tau_{2})$, including 
terms to linear order in $\chi(\tau)$. 
We take the derivative of the self-energy 
with respect to the full Green's function to obtain the vertex function and 
afterwards set the field, $\chi(\tau)$ to zero. The calculation differs from 
standard approaches, in that $\chi(\tau)$ provides a time-dependence so that 
$G_{0}(\tau_{1},\tau_{2})$ depends separately on $\tau_{1}$ and $\tau_{2}$, 
not just on the difference, $(\tau_{1}-\tau_{2})$.

To begin, we introduce an auxiliary Green's function~\cite{brandt_urbanek} 
(as Brandt and Urbanek did in their calculation of the $f$-spectral function)
in the presence of just the charge-coupled field, $\chi$, defined by:
\begin{equation}
g_{aux}(\tau_{1},\tau_{2};\mu) = \frac{ -{\rm Tr_d} T_{\tau} \left<
e^{-\beta \hat{H}_{0}} e^{ \int^{\beta}_{0} {\mathrm d} \overline{\tau} 
\chi(\overline{\tau}) {d}^{\dagger}(\overline{\tau})
{d}(\overline{\tau})} d(\tau_{1}) d^{\dagger}(\tau_{2}) \right> }
{1 + e^{\beta \mu} e^{ \int^{\beta}_{0} {\mathrm d} \overline{\tau} 
\chi(\overline{\tau}){d}^{\dagger}(\overline{\tau})
{d}(\overline{\tau})} },
\label{eq:trace}
\end{equation}
where $ \hat{H}_{0}=-\mu d^{\dagger}d$ and 
we expand the time-dependent field, $\chi$, in a Fourier series of the Bosonic 
frequencies
\begin{equation}
\chi(\tau) = \sum_{l} \chi_{l} e^{i\nu_{l}\tau}.
\end{equation}
Noting that ${d}(\tau) = e^{\tau \hat{H}_{0}}d(0)e^{\tau \hat{H}_{0}}
= e^{-\tau \mu} d(0)$, and taking into account the time ordering, we can easily 
perform the trace in Eq.(\ref{eq:trace}), which yields 
(to linear order in $\chi_{l}$):
\begin{equation}
g_{aux}(\tau_{1},\tau_{2};\mu) = 
e^{\mu(\tau_{1}-\tau_{2})}
\left[1+\sum_{l\neq 0}\frac{\chi_{l}}{i\nu_{l}}
\left(
e^{i\nu_{l}\tau_{1}} - e^{i\nu_{l}\tau_{2}}
\right) \right]
\left[ \frac{- \theta \left( \tau_{1}-\tau_{2} \right) }
{1 + e^{\beta\mu}e^{\beta\chi_{0}}}
+ \frac{\theta \left( \tau_{2}-\tau_{1}\right)}
{1 + e^{-\beta\mu}e^{-\beta\chi_{0}}} \right],
\end{equation}
which depends on both $\tau_1$ and $\tau_2$ (not just $\tau_{1}-\tau_{2}$) 
because of the terms with $\chi_{l\neq 0}$.  Nevertheless, the following 
symmetries are readily proven: 
\begin{eqnarray}
{\mbox {\rm for}} &\tau_{1}& < \tau_{2} < \tau_{1}+\beta \cr
 &g_{aux}&(\tau_{1}+\beta,\tau_{2};\mu) = -g_{aux}(\tau_{1},\tau_{2};\mu) ,
\label{eq: sym1}
\end{eqnarray}
and 
\begin{eqnarray}
{\mbox {\rm for}} &\tau_{2}& < \tau_{1} < \tau_{2}+\beta \cr
 &g_{aux}&(\tau_{1},\tau_{2}+\beta;\mu) = -g_{aux}(\tau_{1},\tau_{2};\mu) .
\label{eq: sym2}
\end{eqnarray}
Hence we can find the Fourier transform of 
$g_{aux}(\tau_{1},\tau_{2};\mu)$ in 
terms of the Fermionic Matsubara frequencies
\begin{equation}
g_{aux}(i\omega_{n},i\omega_{n'};\mu) = 
\int^{\beta}_{0} {\mathrm d}  \tau_{1} \int^{\tau_1}_{-\beta+\tau_1} 
{\mathrm d}  \tau_{2} 
e^{i\omega_{n}\tau_{1}} e^{-i\omega_{n'}\tau_{2}}
g_{aux}(\tau_{1},\tau_{2};\mu).
\end{equation}
The result is that $g_{aux}(i\omega_{n},i\omega_{n'};\mu)$ contains terms that 
are diagonal in $n,n'$ and terms shifted off the diagonal by $l$ that are 
proportional to $\chi_{l}$:
\begin{equation}
g_{aux}(i\omega_{n},i\omega_{n'};\mu) = 
\frac{\delta_{n,n'}}{i\omega_{n}+\mu} + 
\sum_{l} \frac{\delta_{n+l,n'}\chi_{l}}{i\nu_{l}} 
\left( \frac{1}{i\omega_{n'} + \mu} - \frac{1}{i\omega_{n} + \mu} \right) .
\end{equation}

The Green's function for the effective medium of the atomic problem, 
$G_{0}(\tau,\tau')$, is 
obtained as before by including a local time-dependent field, 
$\lambda(\tau - \tau')$, that incorporates 
the effects of propagation on the lattice 
and only depends on the time-difference:
\begin{equation}
G_{0}(\tau,\tau') = \frac{ -{\rm Tr_d} T_{\tau} \left<
e^{-\beta \hat{H}_{0}} e^{ \int^{\beta}_{0} {\mathrm d} \overline{\tau} 
\chi(\overline{\tau}) {d}^{\dagger}(\overline{\tau}) 
{d}(\overline{\tau})} S(\lambda) 
d(\tau_{1}) d^{\dagger}(\tau_{2}) \right> }{Z_{0}(\mu)} .
\label{eq: g0new}
\end{equation}
As the local field only contains diagonal frequency 
components, $\lambda(i\omega_{n})\delta_{nn'}$, the effective-medium Green's 
function is easily obtained from the auxiliary Green's function through:
\begin{equation}
\left[ G_{0} \right]^{-1}_{n,n'} = 
\left[ g_{aux}(\mu) \right]^{-1}_{n,n'} - \lambda(i\omega_{n})\delta_{n,n'} ,
\end{equation}
where the Green's function matrices are inverted in frequency coordinates, 
such that $\left[ G_{0} \right]^{-1}_{n,n'}$ represents the $n,n'$ components of 
the inverse of the matrix $G_{0}(i\omega_{n},i\omega_{n'})$. 
A key aspect of our derivation of the vertex function, 
is that to linear order in a single frequency component of the 
charge-coupled field, $\chi_{l}$, the 
Green's function for the effective medium, $G_{0}(i\omega_{n},i\omega_{n'})$, 
like the auxiliary Green's function, $g_{aux}(i\omega_{n},i\omega_{n'};\mu)$, 
only contains components in the two diagonals given by $\delta_{n,n'}$ and 
$\delta_{n+l,n'}$. 

The full Green's function is defined by:
\begin{equation}
G(\tau_{1},\tau_{2}) = \frac{ -{\rm Tr_{f}Tr_{d}} T_{\tau} \left<
e^{-\beta \hat{H}_{at}} e^{ \int^{\beta}_{0} {\mathrm d} \overline{\tau} 
\chi(\overline{\tau}) {d}^{\dagger}(\overline{\tau})
{d}(\overline{\tau})}  S(\lambda)
d(\tau_{1}) d^{\dagger}(\tau_{2}) \right> }
{Z},
\end{equation} 
where $Z$ is given by Eq.~(\ref{eq: zdef})
and the time dependence of the ${d}$-operators is governed by 
$\hat{H}_{at}$. 
The Green's function contains a trace over $f$-electron states, 
which leads to two terms:
\begin{eqnarray}
G(\tau_{1},\tau_{2}) & = & \frac{ -Tr_{d} T_{\tau} \left<
e^{-\beta \hat{H}_{0}} e^{ \int^{\beta}_{0} {\mathrm d} \overline{\tau} 
\chi(\overline{\tau}) {d}^{\dagger}(\overline{\tau})
{d}(\overline{\tau})} S(\lambda)
d(\tau_{1}) d^{\dagger}(\tau_{2}) \right> }
{Z} \\
& & + e^{-\beta(\epsilon_{f}-\mu)} \frac{ -Tr_{d} T_{\tau} \left<
e^{-\beta \hat{H}_{1}} e^{ \int^{\beta}_{0} {\mathrm d} \overline{\tau} 
\chi(\overline{\tau}) {d}^{\dagger}(\overline{\tau})
{d}(\overline{\tau})}  S(\lambda) 
d(\tau_{1}) d^{\dagger}(\tau_{2}) \right> }{Z} ,
\label{eq:fullG}
\end{eqnarray}
where $\hat{H}_{0}=-\mu d^{\dagger}d$ is the atomic Hamiltonian with no 
$f$-electrons, and $\hat{H}_{1}=(U-\mu) d^{\dagger}d$ is the atomic Hamiltonian 
for $d$-electrons in the presence of one $f$-electron. 

We can now relate the full Green's function to the Green's function of the 
effective medium through the following matrix equation:
\begin{equation}
G(i\omega_{n},i\omega_{n'}) = w_{0} G_{0}(i\omega_{n},i\omega_{n'}) + 
w_{1} \left[ \left(G_{0}^{-1}\right) - U \right]^{-1}_{n,n'}
\label{eq:gaux}
\end{equation}
where $w_{0}=1-w_1$ and $w_{1}$ are given  by Eq.~(\ref{eq: w1def})
and equal the 
fraction of sites with an occupancy of zero and one $f$-electron, respectively. 
In the above, and all following equations, $U$ is the only matrix which is 
necessarily diagonal in frequency space.

Eq.(\ref{eq:gaux}) can be rearranged by multiplying on the left or right by
matrices like $G^{-1}$, $G_0^{-1}$ and $G_0^{-1}-U$
to give the two following matrix equations (with indices suppressed):
\begin{eqnarray}
G_{0}^{-2} - (U+G^{-1}) G_{0}^{-1} + (1-w_{1})UG^{-1}& =& 0,\\
 G_{0}^{-2} - G_{0}^{-1}(U+G^{-1}) + (1-w_{1})UG^{-1}& =& 0.
\end{eqnarray}
Adding these two equations together and collecting terms (noting the 
noncommutativity and hence the ordering 
of the matrices) then yields the following quadratic matrix equation:
\begin{equation}
\left[G_{0}^{-1} -\frac{1}{2}(U+G^{-1})\right]^{2} 
- \frac{1}{4}(U+G^{-1})^{2} + (1-w_{1})UG^{-1} = 0 .
\end{equation}
Substitution for $G_{0}$ by $\Sigma$, using Dyson's equation in the
form
\begin{equation}
\Sigma(i\omega_{n},i\omega_{m}) = \left[G_{0}^{-1}\right]_{n,m}
- \left[G^{-1}\right]_{n,m} ,
\label{eq:self}
\end{equation}
yields:
\begin{equation}
\left[ \Sigma + \frac{1}{2}\left(G^{-1}-U\right)\right]^{2} 
= \frac{1}{4}\left[U^{2} + 2(2w_{1}-1)UG^{-1} + G^{-2}\right],
\label{eq:quad}
\end{equation}
where each term in $G$ and $\Sigma$ is a matrix in frequency-space
and $U$ multiplies the identity matrix. Note that in the limit 
$\chi_{l}=0$, Eq.(\ref{eq:quad}) becomes diagonal, and reduces to the 
conventional quadratic expression for the self-energy in terms of 
$G(i\omega_{n})$ and $w_{1}$ as first determined by Brandt and 
Mielsch~\cite{brandt_mielsch}. 

The irreducible vertex function is defined in frequency space by:
\begin{equation}
\Gamma(i\omega_{n},i\omega_{m};i\omega_{n'},i\omega_{m'}) = 
\int^{\beta}_{0} {\mathrm d} \tau_{1} \int^{\beta+\tau_{1}}_{\tau_{1}} 
{\mathrm d} \tau_{2}
\int^{\beta}_{0} {\mathrm d} \tau_{1'} \int^{\beta+\tau_{1'}}_{\tau_{1'}} 
{\mathrm d} \tau_{2'}
e^{i\omega_{n}\tau_{1}-i\omega_{m}\tau_{2}+
i\omega_{n'}\tau_{1'}-i\omega_{m'}\tau_{2'}}
\Gamma(\tau_{1},\tau_{2};\tau_{1'},\tau_{2'}) .
\end{equation}
As the vertex function is independent of the absolute time, we can change 
variables: $\tau_{1}\mapsto \tau_{1}-\tau_{2'} ,  
\tau_{2}\mapsto \tau_{2}-\tau_{2'} , \tau_{1'}\mapsto \tau_{1'}-\tau_{2'}$ so 
that the $\tau_{2'}$ integral yields a delta function which requires 
$i\omega_{m'} = i\omega_{n} - i\omega_{m} + i\omega_{n'}$. Hence, 
given the difference between two Matsubara frequencies, 
$i\omega_{m}-i\omega_{n} = i\nu_{l}$ is a Bosonic frequency, 
the vertex function becomes
\begin{equation}
\Gamma(i\omega_{n},i\omega_{n'};i\nu_{l}) = \frac{1}{T}
\frac{\delta \Sigma(i\omega_{n},i\omega_{n}+i\nu_{l})}
{\delta G(i\omega_{n'},i\omega_{n'}+i\nu_{l})} .
\end{equation}

The problem is greatly simplified by calculating $\Sigma_{n,m}$ to first order 
in $\chi$, for a single frequency component, $\chi_{l}$. In this case, 
both $\Sigma_{n,m}$ and $G_{n,m}$ contain diagonal terms, and off-diagonal 
terms linear in $\chi_{l}$ where $m=n+l$, so that:
\begin{equation}
\begin{array}{lllll}
\Sigma_{n,m} & = & \Sigma_{n}\delta_{n,m} & + &
 \bar\Sigma_{n}\delta_{n+l,m},\\
G_{n,m} & = & G_{n}\delta_{n,m}  & + & \bar G_{n}\delta_{n+l,m}.
\end{array}
\end{equation}
We are interested in the case where $l \neq 0$, as the zero-frequency response 
(corresponding to a shift of the chemical potential) is 
well-known~\cite{brandt_mielsch,freericks,freericks_zlatic}.

Next, we calculate the diagonal and off-diagonal pieces of the
self-energy. Eq.(\ref{eq:quad}) simplifies from 
a full quadratic matrix equation, to the following coupled equations:
\begin{eqnarray}
\left(\Sigma_{n} - \frac{U}{2} + \frac{1}{2G_{n}}\right)^{2} 
& = & \frac{1}{4G_{n}^{2}} \left\{ 1 +2UG_{n}(2w_{1}-1) + 
U^{2}G_{n}^{2} \right\} \label{eq:self1},\\
\left(\bar\Sigma_{n} - \frac{\bar G_{n}}{2G_{n}G_{n+l}}\right)
\left(\Sigma_{n} +\frac{1}{2G_{n}} +
\Sigma_{n+l} +\frac{1}{2G_{n+l}} - U \right) & = &
\frac{\bar G_{n}}{4G_{n}G_{n+l}}
\left[ 2(1-2w_{1}) U - \frac{ (  G_{n}+G_{n+l})}
{G_{n}G_{n+l}} \right] .
\label{eq:self2}
\end{eqnarray}
Eq.(\ref{eq:self1}) contains only diagonal terms with no dependence on the
finite-frequency field $\chi_{l}$, hence $\Sigma_n$ is unchanged from its
value when $\chi_l=0$. We concentrate on rearranging 
Eq.(\ref{eq:self2}) to obtain the frequency-dependent response. First, 
we multiply both sides of Eq.(\ref{eq:self2}) by 
\begin{equation}
\left(\Sigma_{n} - \frac{U}{2} + \frac{1}{2G_{n}}\right) 
- \left(\Sigma_{n+l} - \frac{U}{2} + \frac{1}{2G_{n+l}}\right) 
\end{equation}
and use Eq.(\ref{eq:self1}) to replace the quadratic terms on the left, giving:
\begin{eqnarray}
\left(\bar\Sigma_{n} - \frac{\bar G_{n}}{2G_{n}G_{n+l}}\right) & &
\left[2U(2w_{1}-1) + \frac{1}{G_{n}} + \frac{1}{G_{n+l}}\right]
\left( \frac{1}{4G_{n}} - \frac{1}{4G_{n+l}}\right) \nonumber\\ 
= & & \frac{\bar G_{n}}{4G_{n}G_{n+l}}
\left[ 2(1-2w_{1}) U - \frac{ (  G_{n}+G_{n+l})}
{G_{n}G_{n+l}} \right]
\left(\Sigma_{n}+\frac{1}{2G_{n}}-\Sigma_{n+l}
-\frac{1}{2G_{n+l}}\right) .
\end{eqnarray}
The above form gives rise to considerable cancellation between the two sides, 
leading to the amazingly simple result:
\begin{equation}
\bar\Sigma_{n} = \bar G_{n} \frac{ \Sigma_{n} - \Sigma_{n+l}}
{G_{n} - G_{n+l} } .
\end{equation}
So the only terms in 
$\Sigma_{n,m}$ that depend on a finite frequency component, $i\nu_{l}$, of the 
$\chi$ field are $\Sigma_{n,n+l}=\bar\Sigma_n$. The only terms which survive 
the limit, 
$\chi_{l} \mapsto 0$ after differentiation, to give the vertex function, are:
\begin{equation}
\Gamma(i\omega_{n},i\omega_{n};i\nu_{l}) = 
\frac{1}{T}\frac{\delta\Sigma( i\omega_{n},i\omega_{n+l} )}
{\delta G( i\omega_{n},i\omega_{n+l} )} 
 =  \frac{1}{T}\frac{\delta\bar\Sigma_{n}}{\delta \bar G_{n}} 
 = \frac{1}{T}\frac{ \Sigma_{n} - \Sigma_{n+l}}
{G_{n} - G_{n+l} } .
\label{eq:BKres}
\end{equation}

The above simple result found from this Baym-Kadanoff analysis is identical
to that found by diagrammatic 
calculations~\cite{shvaika} which give the more complicated form:
\begin{equation}
\Gamma(i\omega_{n},i\omega_{n};i\nu_{l}) = \frac{1}{T}
\frac{w_{0}w_{1}U^{2}}
{\left(1+G_{n}\Sigma_{n}\right)\left[1+G_{n}(\Sigma_{n}-U)\right]
\left(1+G_{n+l}\Sigma_{n+l}\right)\left[1+G_{n+l}(\Sigma_{n+l}-U)\right]
+ w_{0}w_{1}U^{2}G_{n}G_{n+l} } .
\label{eq:diagres}
\end{equation}
The proof of this equivalence is presented in the Appendix.

Since the vertex function is diagonal, we have a simple form for the
dynamical charge susceptibility
\begin{equation}
\chi^{dd}(X;i\nu_l\ne 0)=T\sum_m\frac{\chi_0(X,i\omega_m;i\nu_l)}
{1+\chi_0(X,i\omega_m;i\nu_l)\frac{\Sigma_m-\Sigma_{m+l}}{G_m-G_{m+l}}}.
\label{eq: chi_final}
\end{equation}
In the limit ${\bf q}=0$ ($X=1$), substitution of $\chi_0^{dd}$ from 
Eq.~(\ref{eq: chi0_1}) into Eq.~(\ref{eq: chi_final}) gives
\begin{equation}
\chi^{dd}(1;i\nu_l\ne 0)=-T\sum_m\frac{G_m-G_{m+l}}{i\nu_l}=0,
\label{chi_final_1}
\end{equation}
as expected by our symmetry arguments.  Notice that $\chi_0^{dd}$ does not
vanish; it is the vertex corrections that force the full susceptibility
to vanish.

The final step in our formalism development is to perform the analytic 
continuation of Eq.~(\ref{eq: chi_final}) from the imaginary to the real
axis.  Our method is not completely rigorous, because we are unable
to verify necessary analyticity arguments as described below, but
we compare the direct calculation of the susceptibility on the imaginary
axis to the inferred value there that is found by using the spectral
formula from the real axis.  In nearly
all cases, we have accuracy to better than
one part in 1000, which supports, {\it a posteriori}, that our technique
is valid.

We will consider only the case with $X=-1$ here, since the $X=1$ susceptibility
is trivial and since the $X=0$ case is simpler and can easily be worked out
by following the same steps we use for the chessboard case.
We start by replacing the summation over Matsubara frequencies in 
Eq.~(\ref{eq: chi_final}) by three contour integrals, which are chosen
as in Figure 1 to encircle all of the simple poles at the Fermionic Matsubara
frequencies, and no other poles in the corresponding integrals [in other
words, the only poles contributing are the simple poles from $f(\omega)$ below].
The susceptibility then becomes (for simplicity, we consider $\nu_l>0$ here)
\begin{eqnarray}
\chi^{dd}(-1;i\nu_l)&=&\frac{i}{2\pi}\int_{C_1}d\omega f(\omega)
\frac{ \frac{G_R(\omega)+G_R(\omega+i\nu_l)}{2\omega+2\mu+i\nu_l-
\Sigma_R(\omega)-\Sigma_R(\omega+i\nu_l)} }
{1-\frac{G_R(\omega)+G_R(\omega+i\nu_l)}{2\omega+2\mu+i\nu_l-
\Sigma_R(\omega)-\Sigma_R(\omega+i\nu_l)}\frac{\Sigma_R(\omega)-
\Sigma_R(\omega+i\nu_l)}{G_R(\omega)-G_R(\omega+i\nu_l)}}\cr
&+&
\frac{i}{2\pi}\int_{C_2}d\omega f(\omega)
\frac{ \frac{G_A(\omega)+G_R(\omega+i\nu_l)}{2\omega+2\mu+i\nu_l-
\Sigma_A(\omega)-\Sigma_R(\omega+i\nu_l)} }
{1-\frac{G_A(\omega)+G_R(\omega+i\nu_l)}{2\omega+2\mu+i\nu_l-
\Sigma_A(\omega)-\Sigma_R(\omega+i\nu_l)}\frac{\Sigma_A(\omega)-
\Sigma_R(\omega+i\nu_l)}{G_A(\omega)-G_R(\omega+i\nu_l)}}\cr 
&+&
\frac{i}{2\pi}\int_{C_3}d\omega f(\omega)
\frac{ \frac{G_A(\omega)+G_A(\omega+i\nu_l)}{2\omega+2\mu+i\nu_l-
\Sigma_A(\omega)-\Sigma_A(\omega+i\nu_l)} }
{1-\frac{G_A(\omega)+G_A(\omega+i\nu_l)}{2\omega+2\mu+i\nu_l-
\Sigma_A(\omega)-\Sigma_A(\omega+i\nu_l)}\frac{\Sigma_A(\omega)-
\Sigma_A(\omega+i\nu_l)}{G_A(\omega)-G_A(\omega+i\nu_l)}},
\label{eq: ac1}
\end{eqnarray}
where $f(\omega)=1/[1+\exp(\beta\omega)]$ is the Fermi function and the
subscript $R$ or $A$ refers to the retarded or advanced Green's function
(or self-energy).  The choices of the subscripts are so that the
Green's functions and self-energies are analytic within regions 1, 2,
or 3.  Hence the contours can be deformed until they run parallel to the
real axis, as shown in Figure 2.  We are making an assumption that there
are no other poles present within these regions, when we deform the
contours.  In fact, if we make an analytic continuation of just the
bare susceptibility, then after continuing $i\nu_l$ to $\nu+i\delta$ we do find
a pole that lies in region 2 just below the real axis (at $\omega=-\nu/2-
i\delta)$, but the residue of this
pole vanishes when $\nu$ lies on the real axis.  It is more difficult to
make such an analysis for the full susceptibility, so we rely instead on
the comparison with the direct calculation on the imaginary axis.

When we evaluate the integrals along the lines indicated in Figure 2, we 
will evaluate the Fermi function at $\omega-i\nu_l$, which we set equal
to $f(\omega)$ before continuing the $\nu$ frequency.  Then we can 
evaluate the final result, which becomes
\begin{eqnarray}
\chi^{dd}(-1;\nu)=\frac{i}{2\pi}\int_{-\infty}^{\infty}d\omega\Biggr\{
&f&(\omega)\frac{\frac{G(\omega)+G(\omega+\nu)}{2\omega+2\mu+\nu-
\Sigma(\omega)-\Sigma(\omega+\nu)} }
{1-\frac{G(\omega)+G(\omega+\nu)}{2\omega+2\mu+\nu-
\Sigma(\omega)-\Sigma(\omega+\nu)}\frac{\Sigma(\omega)-
\Sigma(\omega+\nu)}{G(\omega)-G(\omega+\nu)}}\cr
&-&f(\omega-\nu)\frac{\frac{G^*(\omega)+G^*(\omega+\nu)}{2\omega+2\mu+\nu-
\Sigma^*(\omega)-\Sigma^*(\omega+\nu)} }
{1-\frac{G^*(\omega)+G^*(\omega+\nu)}{2\omega+2\mu+\nu-
\Sigma^*(\omega)-\Sigma^*(\omega+\nu)}\frac{\Sigma^*(\omega)-
\Sigma^*(\omega+\nu)}{G^*(\omega)-G^*(\omega+\nu)}}\cr
&-&[f(\omega)-f(\omega-\nu)]
\frac{\frac{G^*(\omega)+G(\omega+\nu)}{2\omega+2\mu+\nu-
\Sigma^*(\omega)-\Sigma(\omega+\nu)} }
{1-\frac{G^*(\omega)+G(\omega+\nu)}{2\omega+2\mu+\nu-
\Sigma^*(\omega)-\Sigma(\omega+\nu)}\frac{\Sigma^*(\omega)-
\Sigma(\omega+\nu)}{G^*(\omega)-G(\omega+\nu)}}\Biggr\},
\label{eq: chi_real_final}
\end{eqnarray}
and we replaced the advanced functions by the complex conjugate of the retarded
Green's function (which is valid on the real axis).

\section{Numerical Results}

We performed a number of different numerical calculations of the charge
response.  To begin, we summarize the $d$-electron spectral functions in the 
weak-coupling $(U=0.25)$ limit, the moderate coupling limit $(U=1)$ and
the strong-coupling limit $(U=4)$, which are illustrated in Figure 3. 
Note that the interacting density of states is temperature independent
for this model (as first shown by Van Dongen\cite{vandongen}). In the
weak-coupling case $(U=0.25)$ the density of states looks Gaussian, with
only small modifications due to the interactions.  At moderate coupling,
$(U=1)$, we find that a pseudogap appears in the density of states at the 
chemical potential.  Note that this is a correlation-induced pseudogap and is
not resulting from the charge-density-wave order of the ground state, since
these results are for the high-temperature homogeneous phase.  Finally, a true
gap develops in strong coupling $(U=4)$, which rapidly approaches $U$ as
$U$ increases in magnitude. In Figure 4
we show plots of the static chessboard susceptibility for the three
different values of $U$ and inset a plot of the transition temperature as
a function of $U/(U+t^*)$.  The chain-dotted
lines are guides to the eye for $T_c$ at the three values of $U$.
The transition temperature has a classic form---increasing for weak-$U$,
reaching a maximum at moderate $U$ and then decreasing in the strong-coupling
regime. Notice how the static
susceptibility diverges as we approach $T_c$ because
we are in the thermodynamic limit.
In Figure 5 we show the dynamical chessboard susceptibility at
five different temperatures ranging from well above $T_c$ to just above $T_c$.  
Notice how there is a temperature
dependence to the charge susceptibility, but how there is no indication of
the phase transition seen in the static susceptibility (recall the 
susceptibility is discontinuous at $\nu=0$ and we are only plotting
the continuous piece of the susceptibility). In Figure 6 we show the same
plot for the local dynamical susceptibility.  Its behavior is quite similar
to that of the chessboard susceptibility, but with a less marked temperature
dependence.  The expected peak in the imaginary part
at low frequencies due to quasiparticle
excitations can be seen, and it has a classic ``triangular'' shape as expected
in a noninteracting system (recall the local susceptibility is the average
over all momentum vectors, so we expect a linear contribution at small frequency
and a curved
drop off at high frequencies as we reach the approximate half bandwidth).
The case of moderate coupling is shown in Figures 7 and 8.  Here,
the chessboard susceptibility actually decreases as $T$ is lowered because 
of the pseudogap in the density of states.  Once again the temperature
dependence of the local susceptibility is much less than the chessboard
susceptibility (and we are including data below the 
$T_c$ to the charge-density-wave instability).  The development from a
single low-energy peak in the real part, to two nearly overlapping peaks
(corresponding to charge transfer excitations), is clear even at moderate
coupling where the density of states has only a pseudogap.  Note once
again that these peak developments arise from correlation effects and not from
an underlying charge-density-wave order because these calculations are
all performed within the homogeneous phase (even when we go below $T_c$). 

In Figures 9 and 10 we show
the results at strong coupling $U=4$. 
The charge susceptibility shows a number of interesting
new features in the strong-coupling regime. The charge-transfer peaks, which are
easiest seen as broad peaks in the imaginary part of the susceptibility,
are present at the expected locations of $\pm U$.  At high temperatures, the
system has a large peak in the real part of the susceptibility near
zero frequency, which rapidly decreases as $T$ is lowered, and becomes 
unnoticable at low temperatures.  The imaginary part goes from having
a linear behavior at low frequency to being essentially zero.  Hence,
in addition to the energy scale of the order of $U$ (corresponding to the
large charge-transfer peaks in the imaginary part of the susceptibility) there 
is a low-energy scale on the order of $t^{*2}/U$ that determines the 
low-temperature evolution of the charge excitations and the energy scale
of those low-energy excitations. The high-energy features are ``frozen in''
at a high temperature scale, and there is strong temperature evolution in the 
low-frequency regime until $T$ becomes smaller than the low-energy scale
where the low-energy features are ``frozen in.'' We believe these low-energy
charge excitations are associated with virtual hopping processes, where
an electron virtually hops onto a site occupied by a localized electron, and 
then hops off that site. Such processes turn off at temperatures below 
$t^{*2}/U$.  Once again the dependence of the chessboard
and the local susceptibilities are similar and there is no signal of the phase
transition in the dynamical piece of the charge susceptibility.  
We find that we lose some accuracy
in our calculations at strong coupling, perhaps arising from the gap in the
density of states and its effect on the low-temperature charge dynamics.
Another possibility is that there are additional poles and resides that need to
be taken into account in our ``approximate''
analytic continuation.  One illustration of the
numerical difficulties is that we no longer have an accuracy of one part in
a thousand (it is typically a few parts in a thousand)
when we evaluate the charge susceptibility on the imaginary
axis by using the spectral formula and comparing with a direct calculation
on the imaginary axis.  Another illustration is that at the lowest temperatures
the imaginary part of the spectral function becomes slightly negative for
a frequency range in the vicinity of $0.5<\nu<2$ (where the imaginary part
is approximately $-0.0003$), which must be a numerical artifact due to the
fact that the spectral function is known to be
nonnegative for positive values of the frequency.  

\section{Conclusions}

We have presented a nontrivial example of the Baym-Kadanoff formalism
to derive the dynamical susceptibilities of the Falicov-Kimball model.
The dynamical susceptibility breaks into two pieces, a static piece that
reflects the coupling of the system to the charge-density-wave distortions, and
a frequency-dependent piece, which is less affected by any underlying 
charge-density-wave order. We developed the formalism to evaluate this
frequency-dependent piece of the susceptibility exactly on the imaginary
axis.  We also developed an analytic-continuation scheme that we were unable
to establish with full rigor due to the possibility of a small contribution
from neglected residues of poles that could arise within the 
analytic-continuation procedure, but which is nevertheless quite accurate
over a wide range of parameters.

We find a number of interesting properties of this model: (1) the 
static and dynamical susceptibilities are decoupled---the static susceptibility
diverges at $T_c$, but no signal of this phase transition is seen in the
dynamical susceptibility; (2) the momentum dependence of the dynamical
susceptibility is not too strong around the chessboard point---we see little
variation between the $X=-1$ and $X=0$ susceptibilities, but the $X=1$
susceptibility vanishes by symmetry, so momentum dependence is stronger near
the Brillouin zone center; and (3) the dynamical susceptibilities do
possess temperature dependence even though the interacting density of states
is temperature independent.  The temperature dependence is most striking 
in the strong-coupling regime where a low-energy scale on the order of 
$t^{*2}/U$ arises and the charge susceptibility has a strong temperature
dependence near this approximate temperature.

Our application of the Baym-Kadanoff formalism
provides an interesting counterpoint to the diagrammatic derivation about the
atomic limit, and can be applied to other related problems where the
so-called ``static approximation'' is exact.  We have not examined any
effects away from half-filling or at incommensurate wavevectors here.  We leave
those tasks to another publication.

\acknowledgments

This research was supported by the Office of Naval Research under grant 
number N00014-99-1-0328.  We acknowledge useful discussions with T. Devereaux,
D. Hess, J. Serene, and A. Shvaika.

\appendix

\section{Equivalence to the atomic diagrammatic approach}

To compare our result with the diagrammatic approach~\cite{shvaika}, we 
must begin with Shvaika's result for the frequency-dependent piece
of the reducible vertex function, 
\begin{eqnarray}
\widetilde{\Gamma}(&i\omega_{n}&,i\omega_{n'};i\nu_{l}\neq 0) = \cr
&~&\frac{1}{T}\frac{\delta_{nn'}}
{G_{n}G_{n'}G_{n+l}G_{n'+l}}
\frac{U^{2}w_{1}(1-w_{1})}{\left(i\omega_{n}+\mu -\lambda_{n}\right)
\left(i\omega_{n}+\mu -\lambda_{n}-U\right)
\left(i\omega_{n+l}+\mu -\lambda_{n+l}\right)
\left(i\omega_{n+l}+\mu -\lambda_{n+l}-U\right) },
\label{eq:Shvaika}
\end{eqnarray}
By making the replacement, 
$i\omega_{n}+\mu -\lambda_{n} = \left[G_{0}(i\omega_{n})\right]^{-1} 
= \left(G_{n}\right)^{-1} + \Sigma_{n}$, Eq.(\ref{eq:Shvaika}) 
is equivalent to:
\begin{equation}
\frac{1}{T}\frac{\delta_{nn'} w_{0}w_{1}U^{2}}
{\left(1+G_{n}\Sigma_{n}\right)\left[1+G_{n}(\Sigma_{n}-U)\right]
\left(1+G_{n+l}\Sigma_{n+l}\right)\left[1+G_{n+l}(\Sigma_{n+l}-U)\right] } .
\end{equation}
Solving Dyson's equation for the irreducible vertex function, 
$\Gamma(i\omega_{n},i\omega_{n'};i\nu_{l}\ne 0)$:
\begin{equation}
\widetilde{\Gamma}(i\omega_{n},i\omega_{n'};i\nu_{l}\ne 0) = 
\Gamma(i\omega_{n},i\omega_{n'};i\nu_{l}\ne 0) - 
T\sum_{n''} \Gamma(i\omega_{n},i\omega_{n''};i\nu_{l}\ne 0) G_{n''} G_{n''+l}
\widetilde{\Gamma}(i\omega_{n''},i\omega_{n'};i\nu_{l}\ne 0),
\end{equation}
then yields Eq.~(\ref{eq:diagres}). 

In order to prove that our simple result, Eq.(\ref{eq:BKres}) 
and the diagrammatic result, Eq.(\ref{eq:diagres}) are equivalent 
to each other, we first demonstrate the existence of a useful identity. 
Eq.(\ref{eq:self1}) can be expanded and multiplied by $G_{n}$ to yield:
\begin{eqnarray}
w_{1}U & = & \Sigma_{n}\left(1-UG_{n}+\Sigma_{n}G_{n}\right), \label{eq:w1}\\
& = & \Sigma_{n+l}\left(1-UG_{n+l}+\Sigma_{n+l}G_{n+l}\right), \label{eq:w2}
\end{eqnarray}
where the second equation is evaluated equivalently, with $n \mapsto n+l$. 
Eq.~(\ref{eq:w1}) is used twice to simplify the product given below:
\begin{eqnarray}
\left(1+G_{n}\Sigma_{n}\right)\left[1+G_{n}(\Sigma_{n}-U)\right]
& = &  1 + G_{n}\Sigma_{n} - UG_{n} +w_{1}U,\\
& = & w_{1}U\left(\frac{1}{\Sigma_{n}} + G_{n}\right) ,
\end{eqnarray}
and similarly for $n \mapsto n+l$.
 
Hence the expression on the right of Eq.~(\ref{eq:diagres}) is rewritten as:
\begin{equation}
\frac{1}{T}
\frac{w_{0}w_{1}U^{2}}{w_{1}^{2}U^{2}\left(\frac{1}{\Sigma_{n}} + G_{n}\right)
\left(\frac{1}{\Sigma_{n+l}} + G_{n+l}\right) +G_{n}G_{n+l}w_{0}w_{1}U^{2} },
\end{equation} 
and multiplication of top and bottom by 
$(\Sigma_{n}-\Sigma_{n+l})\Sigma_{n}\Sigma_{n+l}$ leads to:
\begin{equation}
\frac{1}{T}
\frac{(\Sigma_{n}-\Sigma_{n+l})\Sigma_{n}\Sigma_{n+l}(1-w_{1})}
{w_{1}\left[\Sigma_{n}-\Sigma_{n+l} + G_{n}\Sigma_{n}(\Sigma_{n}-\Sigma_{n+l}) 
+ G_{n+l}\Sigma_{n+l}(\Sigma_{n}-\Sigma_{n+l})\right] 
+ G_{n}G_{n+l}\Sigma_{n}\Sigma_{n+l}(\Sigma_{n}-\Sigma_{n+l})} .
\end{equation}
Now Eqs.~(\ref{eq:w1}-\ref{eq:w2}) are further used to make the replacements 
$G_{n}\Sigma_{n}^{2} \mapsto w_{1}U - \Sigma_{n} +UG_{n}\Sigma_{n}$ and 
$G_{n+l}\Sigma_{n+l}^{2} \mapsto w_{1}U - \Sigma_{n+l} +UG_{n+l}\Sigma_{n+l}$ 
for all quadratic terms in the denominator. Numerous cancellations then lead 
to:
\begin{eqnarray}
\Gamma(i\omega_{n},i\omega_{n};i\nu_{l}\ne 0) & = &\frac{1}{T}
\frac{(\Sigma_{n}-\Sigma_{n+l})\Sigma_{n}\Sigma_{n+l}(1-w_{1})}
{w_{1}\left(\Sigma_{n+l}\Sigma_{n}G_{n+l} - \Sigma_{n+l}\Sigma_{n}G_{n}\right) 
- \Sigma_{n+l}\Sigma_{n}G_{n+l} + \Sigma_{n+l}\Sigma_{n}G_{n} },\\
& = & \frac{1}{T}\frac{\Sigma_{n}-\Sigma_{n+l}}{G_{n}-G_{n+l}} ,
\end{eqnarray} 
which yields our final result, Eq.~(\ref{eq:BKres}).

\begin{figure}
\epsfxsize=2.5in
\epsffile{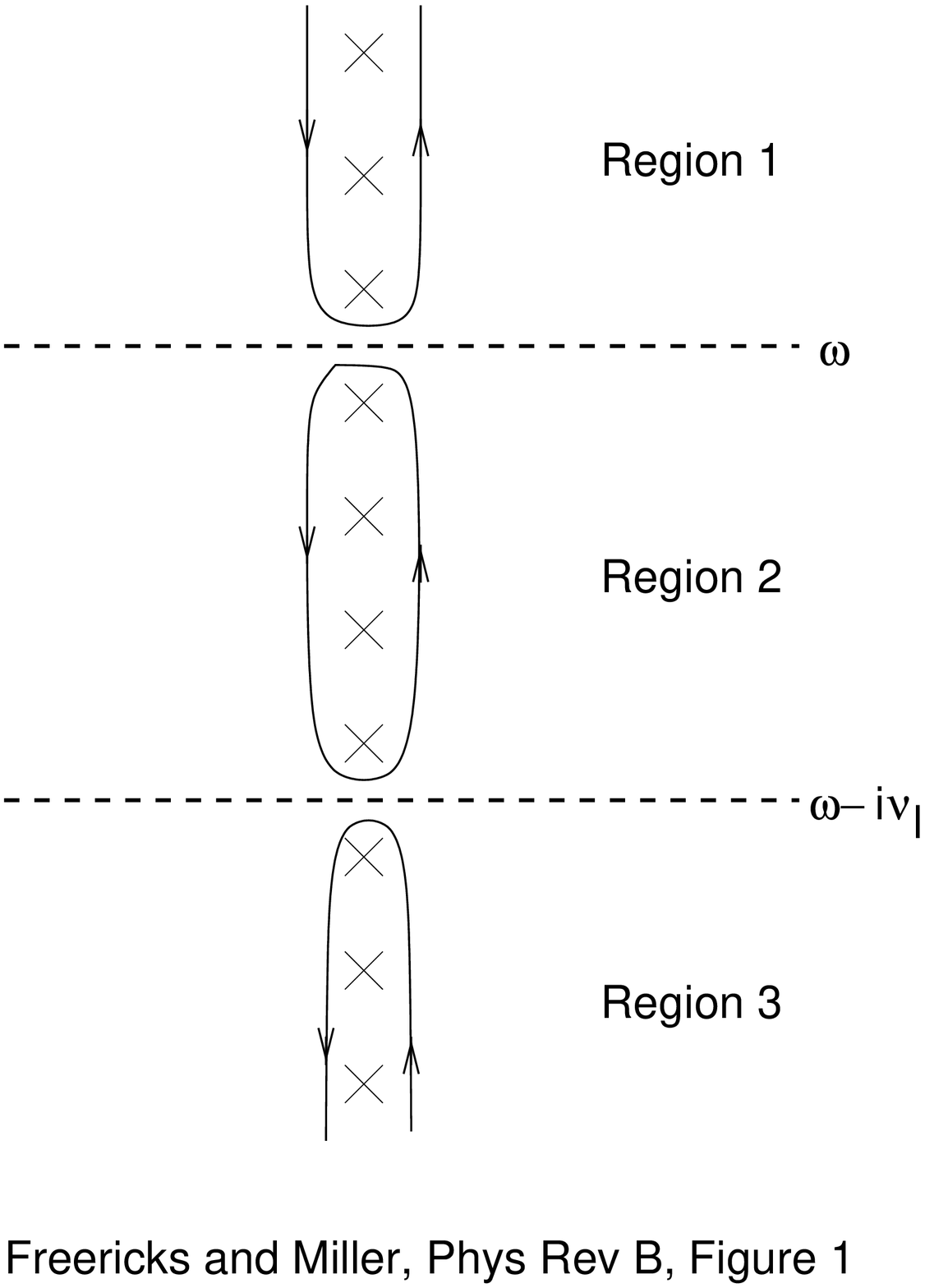}
\caption{Contour integral for evaluating the Matsubara frequency summations 
of the charge susceptibility.  The x's mark the locations of Fermionic
Matsubara frequencies.  The contours enclose all Matsubara frequencies,
but no other poles of the system.  Note that we divide the complex plane
into three regions:
(i) region 1, where the imaginary part is greater than zero; (ii) region 2,
where the imaginary part lies between zero and $-i\nu_l$; and (iii) region 3,
where the imaginary part is less than $-i\nu_l$.}
\end{figure}

\begin{figure}
\epsfxsize=2.5in
\epsffile{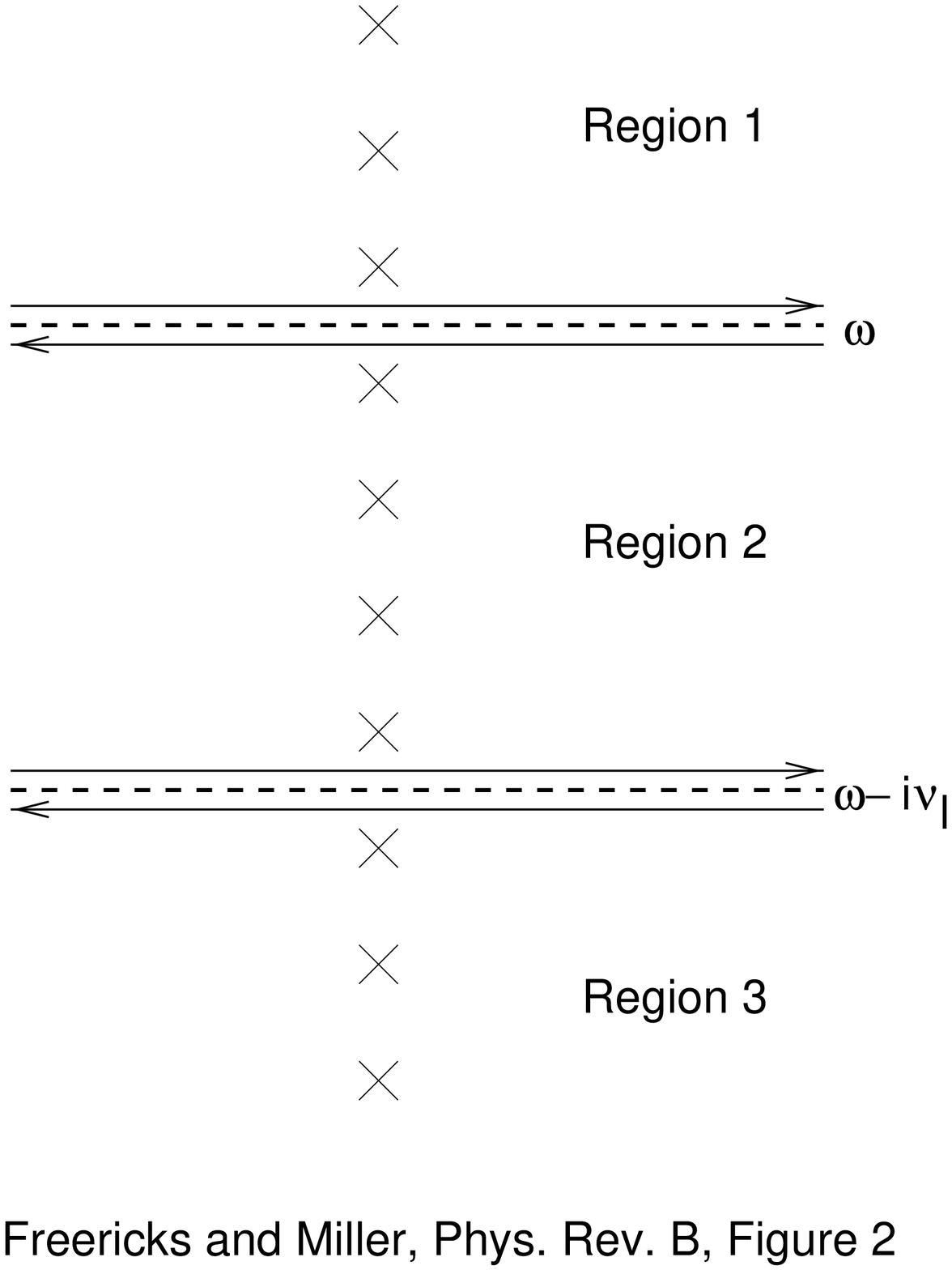}
\caption{Deformation of the contours needed for evaluation of the susceptibility
on the real axis.  The integrations are parallel to the real axis.}
\end{figure}

\begin{figure}
\epsfxsize=3.5in
\epsffile{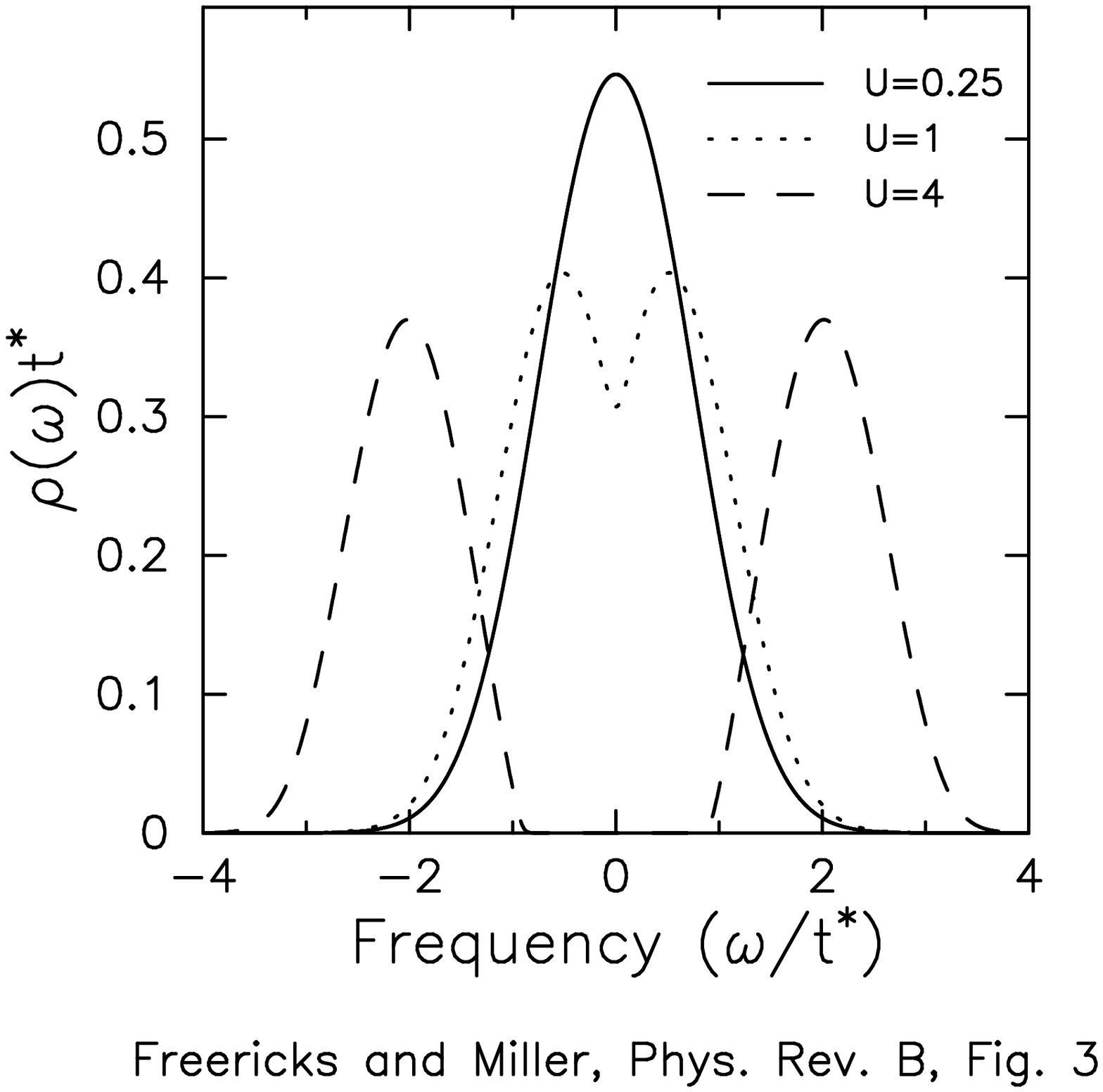}
\caption{Interacting density of states for the spinless Falicov-Kimball
model at half filling and $U=0.25$ (solid line), $U=1$ (dotted line),
$U=4$ (dashed line).  Note how the density of states evolves from being
essentially Gaussian to developing a pseudogap and then a real gap as the
interaction strength is increased.  The density of states is temperature
independent for this model.}
\end{figure}

\begin{figure}
\epsfxsize=3.5in
\epsffile{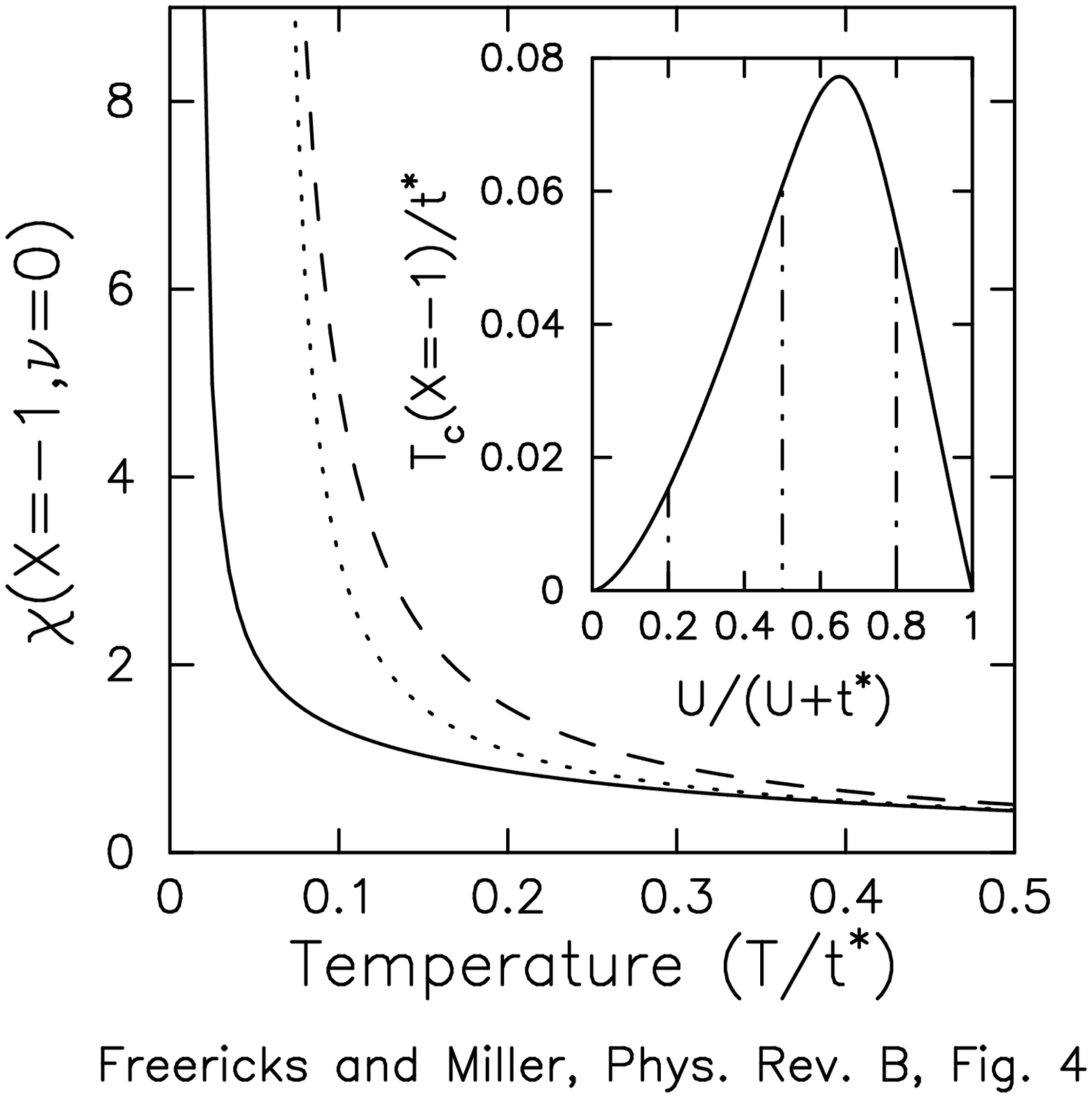}
\caption{Static charge susceptibility for the chessboard charge-density-wave
plotted versus temperature for three values of $U$: (i) $U=0.25$, solid line,
$T_c=0.0153$; (ii) $U=1$, dashed line, $T_c=0.0608$; and (iii) $U=4$,
dotted line, $T_c=0.0547$.  Inset is the charge-density-wave transition 
temperature plotted as a function of the interaction strength $U/(U+t^*)$.}
\end{figure}

\begin{figure}
\epsfxsize=3.5in
\epsffile{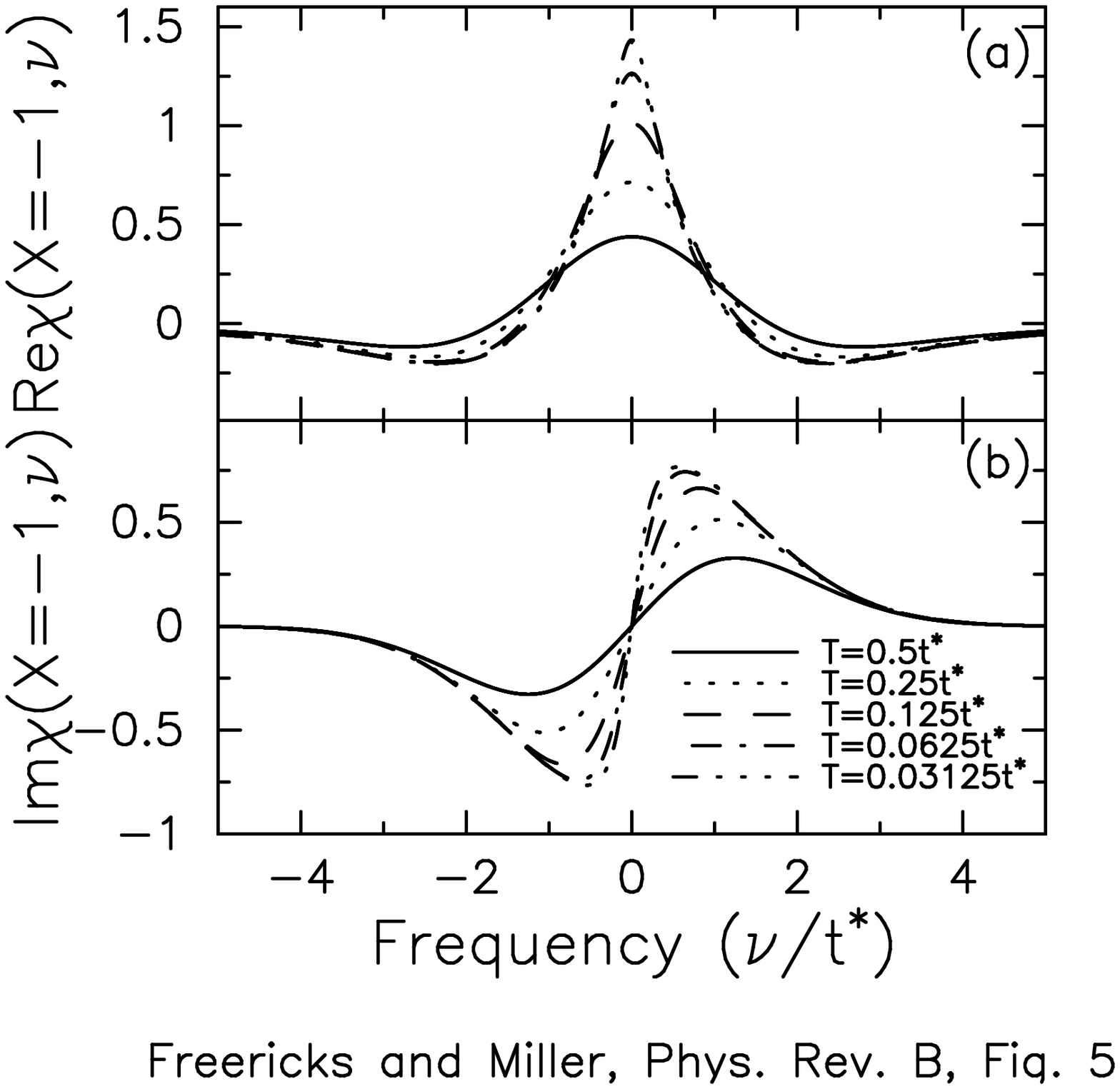}
\caption{Dynamical chessboard charge-density-wave susceptibility as a
function of temperature for the case $U=0.25$.  The real part is plotted
in (a) and the imaginary part in (b).  Five different temperatures are shown:
$T=0.5$ (solid line), $T=0.25$ (dotted line), $T=0.125$ (dashed line),
$T=0.0625$ (chain-dotted line) and $T=0.03125$ (chain-double-dotted line). The
real part of the susceptibility becomes more sharply peaked at zero
frequency, but shows no sign of diverging (because of the decoupling of
the static and dynamic susceptibilities).  The imaginary part sharpens
as $T$ is lowered, and has a shape reminiscent of that of a noninteracting
system arising from the quasiparticle excitations.}
\end{figure}

\begin{figure}
\epsfxsize=3.5in
\epsffile{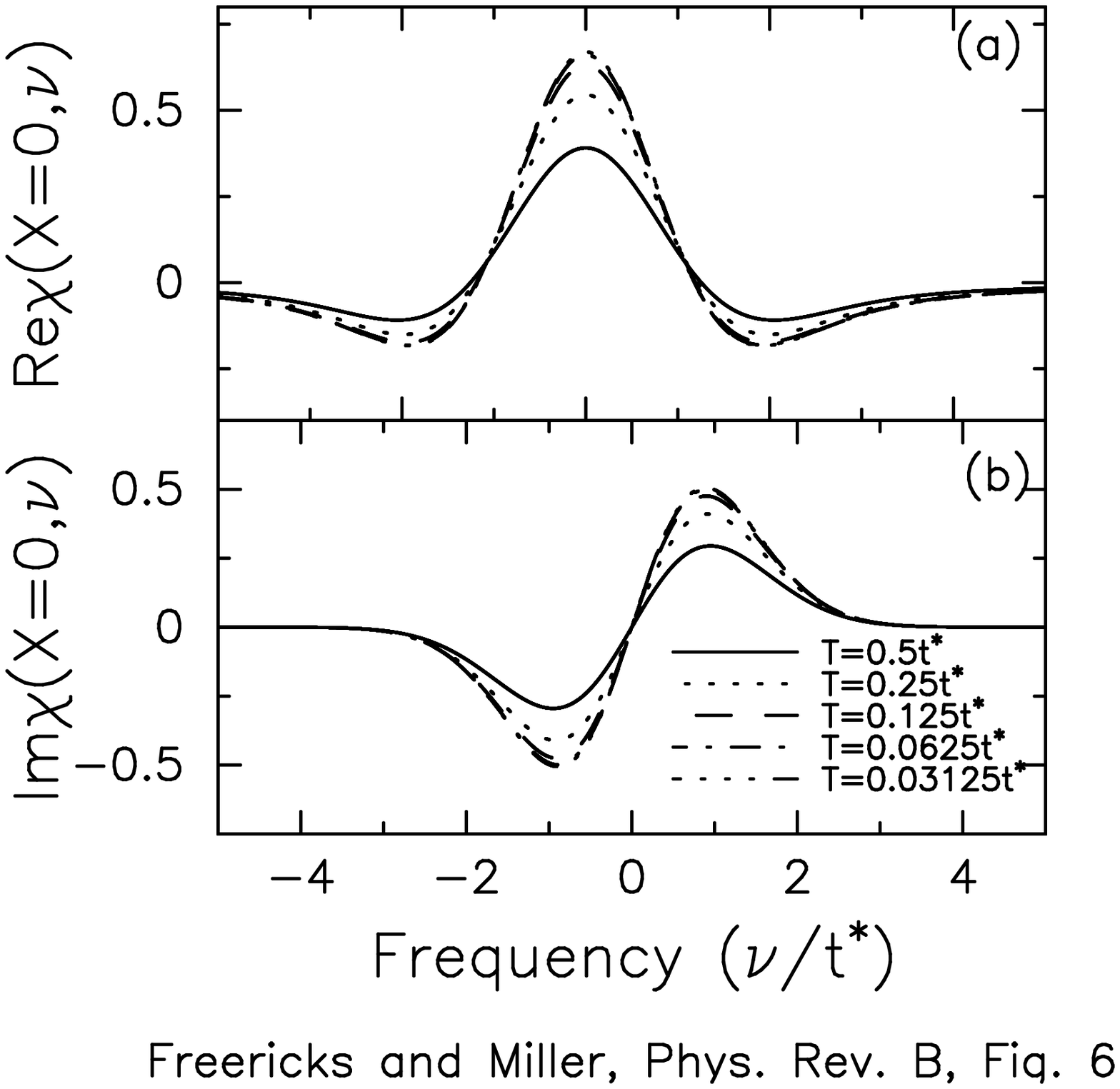}
\caption{Dynamical local charge-density-wave susceptibility as a
function of temperature for the case $U=0.25$.  The real part is plotted
in (a) and the imaginary part in (b).  Five different temperatures are shown:
$T=0.5$ (solid line), $T=0.25$ (dotted line), $T=0.125$ (dashed line),
$T=0.0625$ (chain-dotted line) and $T=0.03125$ (chain-double-dotted line). The
form of the susceptibility is similar to the chessboard case, except the
system reaches the low-temperature limit much more rapidly, and the imaginary
part is pushed somewhat farther from zero frequency.}
\end{figure}

\begin{figure}
\epsfxsize=3.5in
\epsffile{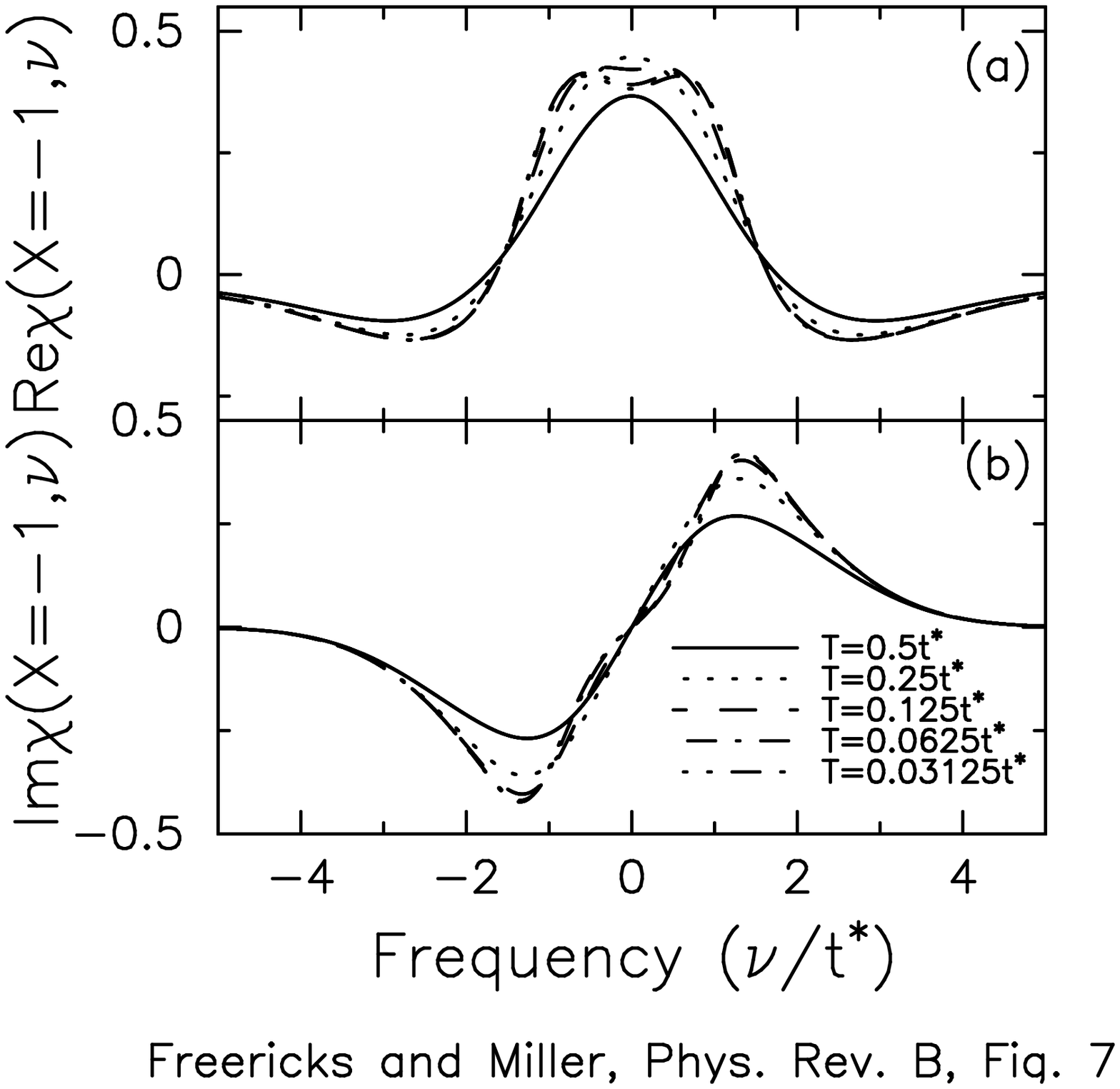}
\caption{Dynamical chessboard charge-density-wave susceptibility as a
function of temperature for the case $U=1$.  The real part is plotted
in (a) and the imaginary part in (b).  Five different temperatures are shown:
$T=0.5$ (solid line), $T=0.25$ (dotted line), $T=0.125$ (dashed line),
$T=0.0625$ (chain-dotted line) and $T=0.03125$ (chain-double-dotted line). The
real part of the susceptibility initially increases as $T$ is lowered at
low frequencies, but then decreases and starts to develop additional peaks
as the effects of the pseudogap are felt.}
\end{figure}

\begin{figure}
\epsfxsize=3.5in
\epsffile{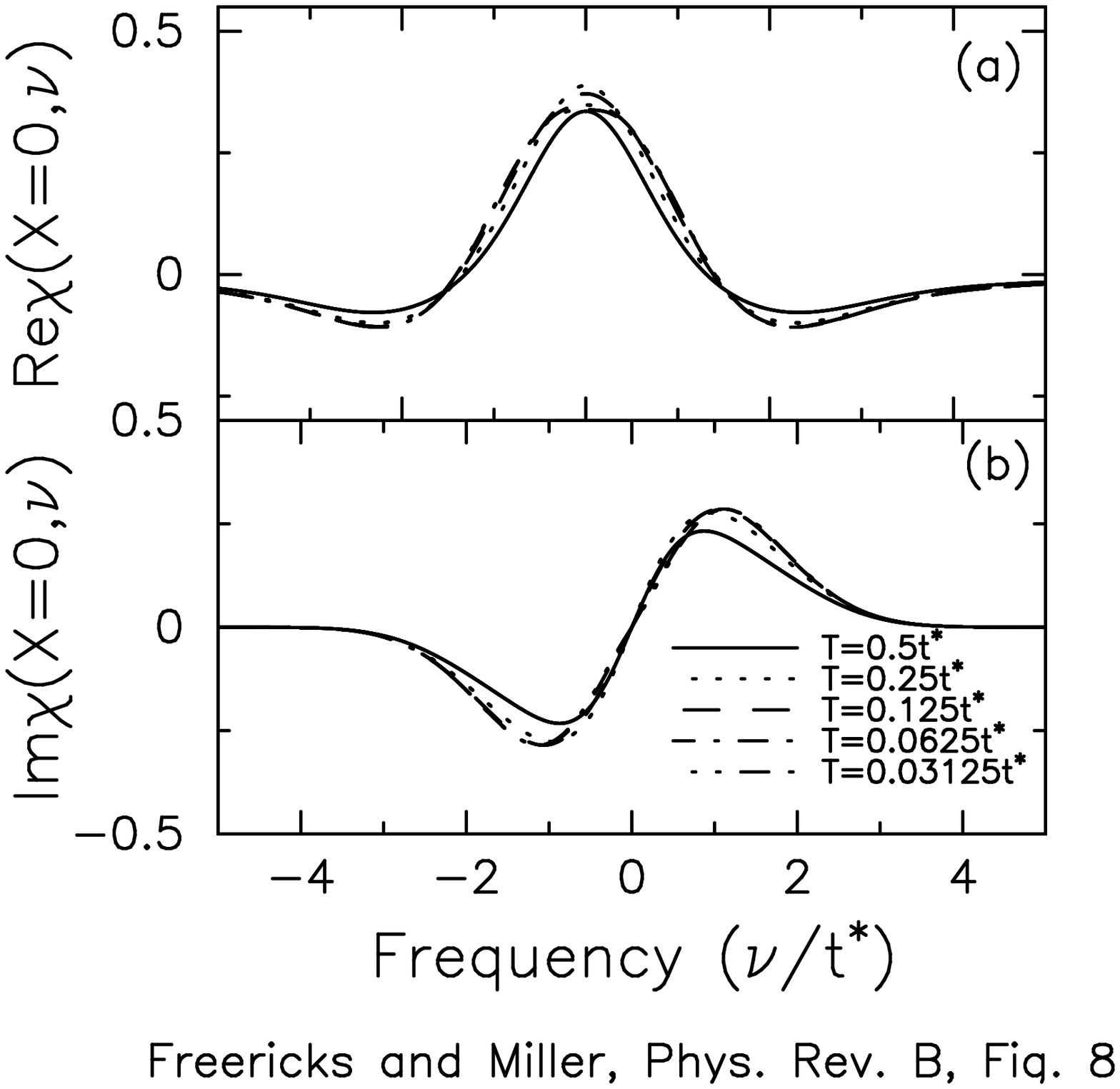}
\caption{Dynamical local charge-density-wave susceptibility as a
function of temperature for the case $U=1$.  The real part is plotted
in (a) and the imaginary part in (b).  Five different temperatures are shown:
$T=0.5$ (solid line), $T=0.25$ (dotted line), $T=0.125$ (dashed line),
$T=0.0625$ (chain-dotted line) and $T=0.03125$ (chain-double-dotted line). We
once again see a similarity with the chessboard phase, but the temperature
dependence is much reduced here.}
\end{figure}

\begin{figure}
\epsfxsize=3.5in
\epsffile{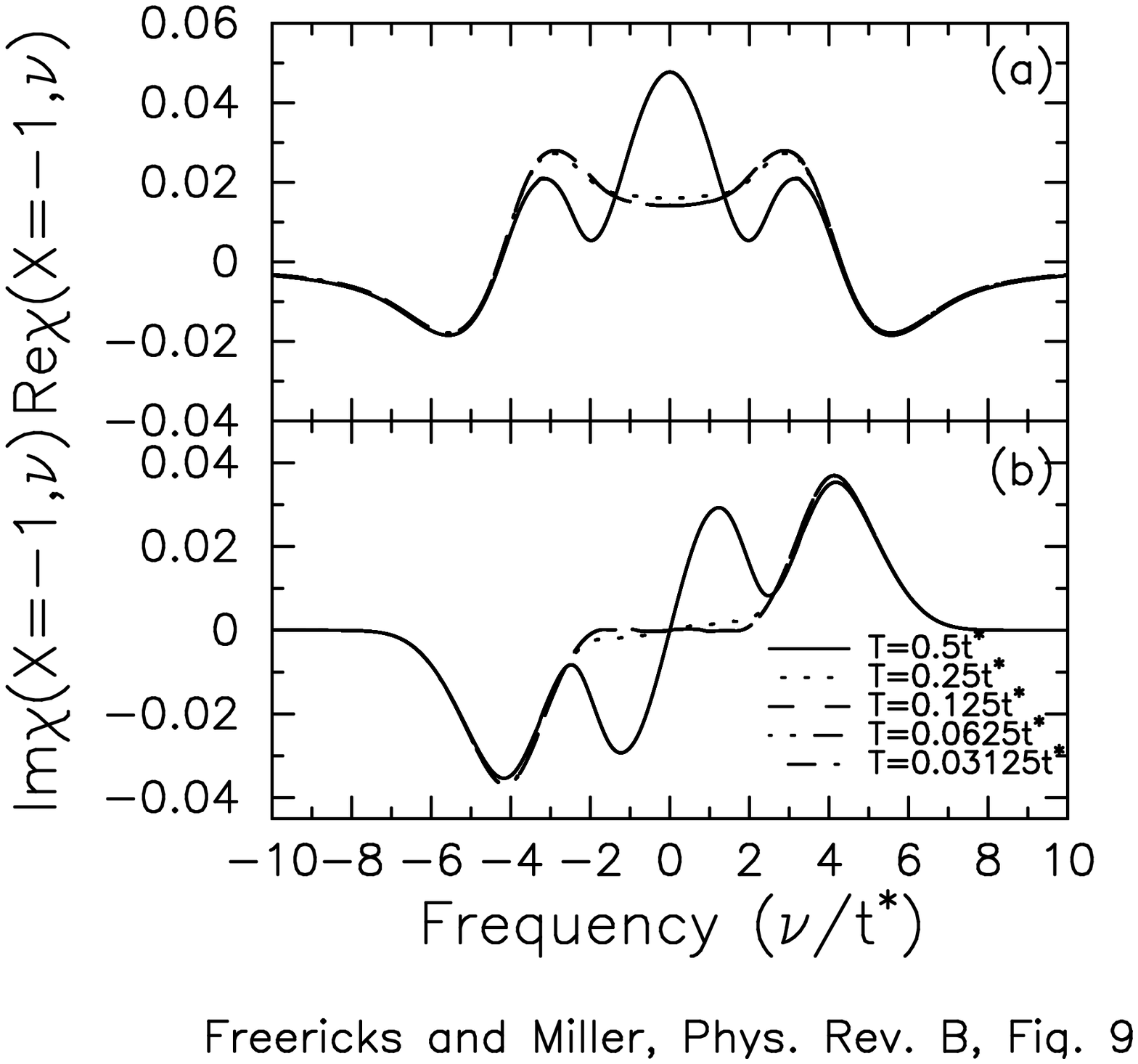}
\caption{Dynamical chessboard charge-density-wave susceptibility as a
function of temperature for the case $U=4$.  The real part is plotted
in (a) and the imaginary part in (b).  Five different temperatures are shown:
$T=0.5$ (solid line), $T=0.25$ (dotted line), $T=0.125$ (dashed line),
$T=0.0625$ (chain-dotted line) and $T=0.03125$ (chain-double-dotted line).
Note how a substantial low-energy feature, present at $T=0.5$ disappears rapidly
as $T$ is lowered.  The temperature scale for this evolution is on
the order of $T\approx t^{*2}/U=0.25$.  For temperatures below this energy
scale the system's behavior is governed most strongly by the energy gap
in the density of states.  We find that the imaginary part of the charge
susceptibility goes slightly negative over a frequency range around 
$0.5<\nu<2$, which we believe to be an artifact of the numerical analytic
continuation procedure in the presence of a gap.} 
\end{figure}

\begin{figure}
\epsfxsize=3.5in
\epsffile{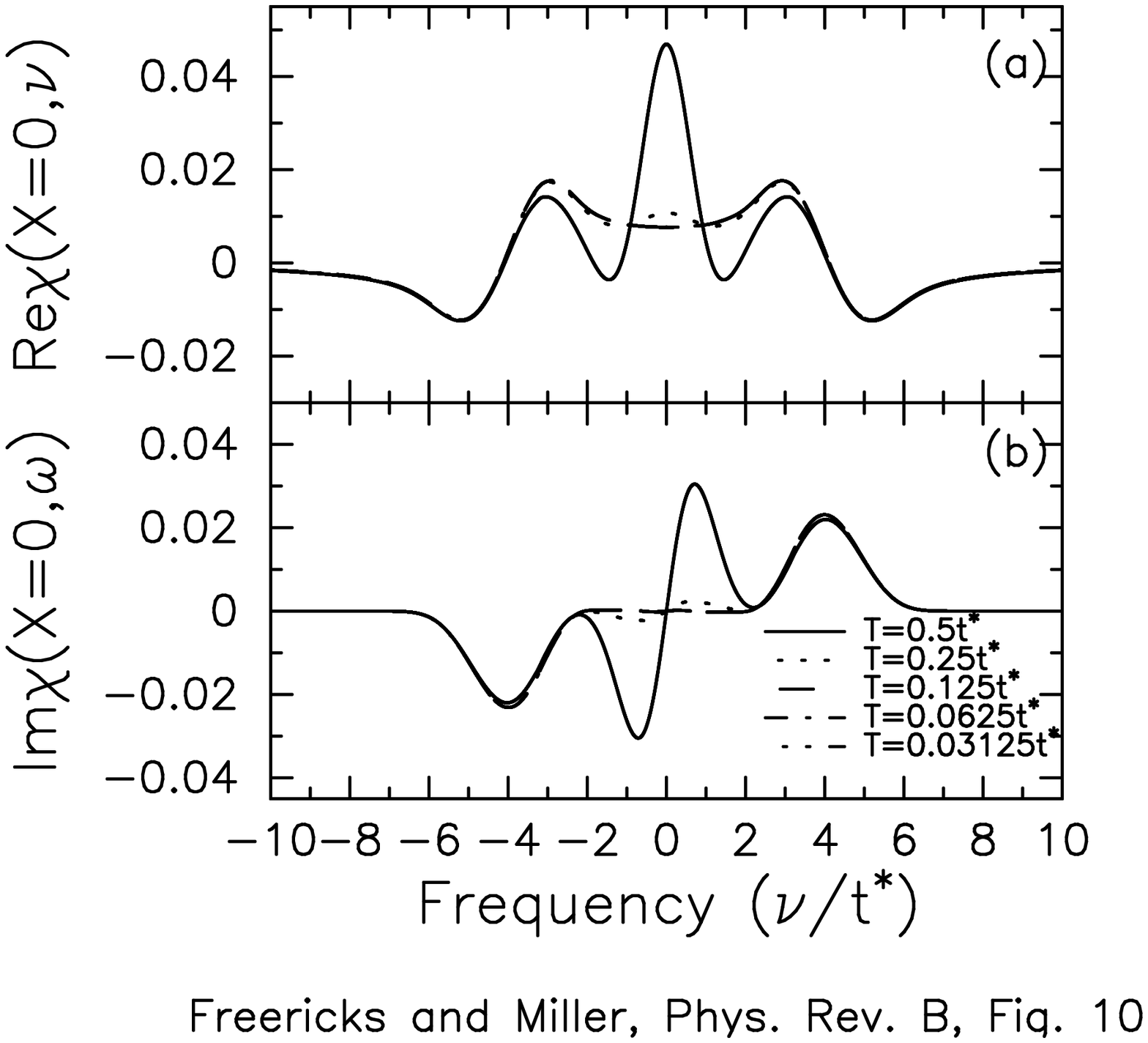}
\caption{Dynamical local charge-density-wave susceptibility as a
function of temperature for the case $U=4$.  The real part is plotted
in (a) and the imaginary part in (b).  Five different temperatures are shown:
$T=0.5$ (solid line), $T=0.25$ (dotted line), $T=0.125$ (dashed line),
$T=0.0625$ (chain-dotted line) and $T=0.03125$ (chain-double-dotted line). The
results here are similar to those for the chessboard susceptibility, except
here we have a slower temperature dependence, as a $\nu=0$ peak in the
real part of the charge susceptibility can still be seen at $T=0.25$, and it
has already vanished in the chessboard case.}
\end{figure}


\begin{thebibliography}{99}
\bibitem{falicov_kimball} L. M. Falicov and J. C. Kimball, Phys. Rev. Lett.
{\bf 22}, 997 (1969); R. Ramirez, L. M. Falicov, and J. C. Kimball, Phys.
Rev. B {\bf 2}, 3383 (1970).
\bibitem{brandt_mielsch} U. Brandt and C. Mielsch, Z. Phys. B {\bf 75},
365 (1989); {\bf 79}, 295 (1990). 
\bibitem{freericks} J. K. Freericks, Phys. Rev. B {\bf 47}, 9263 (1993); 
J. K. Freericks and R. Lemanski, Phys. Rev. B {\bf 62}, XXX (2000).
\bibitem{shvaika} A. M. Shvaika, Physica C (to be published) and
private communication.   
\bibitem{freericks_zlatic}  J. K. Freericks and V. Zlat\'ic, Phys. Rev. B
{\bf 58}, 322 (1998).
\bibitem{mueller-hartmann} E. M\"uller-Hartmann, Z. Phys. B {\bf 74}, 507 
(1989); {\bf 76}, 211 (1989).
\bibitem{Baym1} G. Baym and L. P. Kadanoff, Phys. Rev., {\bf 124}, 287 (1961).
\bibitem{Baym2} G. Baym, Phys. Rev. {\bf 127}, 1391 (1962).
\bibitem{brandt_urbanek} U. Brandt and M. P. Urbanek, Z. Phys. B {\bf 89}, 
297 (1992).
\bibitem{vandongen} P. G. J. Van Dongen, Phys. Rev. B {\bf 45}, 2267 (1992).
\end{thebibliography}
\end{document}